\documentclass[%
 reprint,
showpacs,preprintnumbers,
 amsmath,amssymb,
 aps,
pra,
longbibliography
]{revtex4-1}

\usepackage[utf8]{inputenc}	%Tillader danske tegn
\usepackage[T1]{fontenc}	%Tillader danske tegn
\usepackage{graphicx}		%Tillader indsættelse af billeder
\usepackage{color}

\usepackage{hyperref}
\usepackage{url}

\usepackage{braket}
\usepackage[position=top,caption=false]{subfig}

\usepackage {natbib}

\let\originaleqref\eqref
\renewcommand{\eqref}{Eq.~\originaleqref}

\newcommand{\fref}[1]{\figurename~\ref{#1}}
\newcommand{\Tr}[1]{\mathrm{Tr}\left({#1}\right)}

\newcommand{\e}[1]{\mathrm{e}^{#1}}
\graphicspath{{/home/alexander/Dropbox/PhD/Homodyne detection/Billeder/}{subdir3/}}

\begin{document}
\title{Quantum Zeno effect in parameter estimation}
\author{Alexander Holm Kiilerich}
\email{kiilerich@phys.au.dk}
\author{Klaus Mølmer}

\date{\today}
\affiliation{Department of Physics and Astronomy, Aarhus University, Ny Munkegade 120, DK 8000 Aarhus C. Denmark}
\date{\today}

\bigskip

\begin{abstract} %skriv abstract her:
The quantum Zeno effect freezes the evolution of a quantum system subject to frequent measurements. We apply a Fisher information analysis to show that because of this effect, a closed quantum system should be probed as rarely as possible while a dissipative quantum systems should be probed at specifically determined intervals to yield the optimal estimation of parameters governing the system dynamics. With a Bayesian analysis we show that a few frequent measurements are needed to identify the parameter region within which the Fisher information analysis applies.

\end{abstract}

\maketitle
\noindent

\section{Introduction}
Quantum systems, ranging from atomic systems to field modes and mechanical devices are useful precision probes for a variety of physical properties and phenomena \cite{PhysRevLett.96.010401,Giovannetti19112004}. When we deal with a single quantum system, we must take into account that the measurements by which we extract information about its evolution yield random results and that they cause a measurement back action on the system. For some problems this back action may be favorable as it randomly quenches the system and thus triggers a transient evolution with temporal signal correlations and these may depend more strongly than the steady state on the desired physical properties \cite{PhysRevA.89.052110,PhysRevA.91.012119}.

For too strong or too frequent measurements, however, the back action may completely dominate the evolution. This is manifested in the Quantum Zeno effect (QZE) \cite{ZenoNamed}, named after Zeno's arrow paradox, which inhibits population transfer between discrete states in a quantum system due to frequent observations. The effect was first illustrated in ion trap experiments \cite{PhysRevA.41.2295} and has since been used to account for experiments on many other systems, e.g \cite{BoseImplementation,PhysRevLett.87.040402}.

The intuition behind the quantum Zeno effect has stimulated proposals for application in quantum information processing \cite{perspectives}, e.g. for entanglement protection \cite{PhysRevLett.100.090503}, preservation of systems in decoherence free subspaces \cite{BoseImplementation,PhysRevLett.85.1762} and error suppression in quantum computing \cite{PhysRevA.70.062302}. In many of these studies the measurement back action is actually not unambiguously identified as the crucial element, but experiments and calculations have clearly established the intuition behind the quantum Zeno effect as a useful and inspirational source for new ideas and proposals.

In this study, we focus on information retrieved directly by the measurements and, closer to the original Zeno paradox, we ask to what extent frequent measurements on a quantum system prevent the observation - through the same measurements - of its dynamical evolution. This analysis is pertinent for the use of quantum systems as sensitive probes, and we shall quantify the information content in measurement records and derive theoretical expressions for the Fisher information \cite{Fisher}, to reveal the extent to which high precision probing is inhibited by the quantum Zeno effect in the limit of strong frequent measurements.
The Fisher information quantifies the asymptotic precision in any estimate and has in recent works been applied to analyze quantum parameter estimation in various schemes of repetitive and continuous measurements, see e.g. \cite{PhysRevA.89.052110,PhysRevLett.112.170401,PhysRevA.83.062324,Asymptotic,Ferrie2013,PhysRevA.91.012119}.

\begin{figure}
\centering
\includegraphics[trim=0 0 0 0,width=0.95\columnwidth]{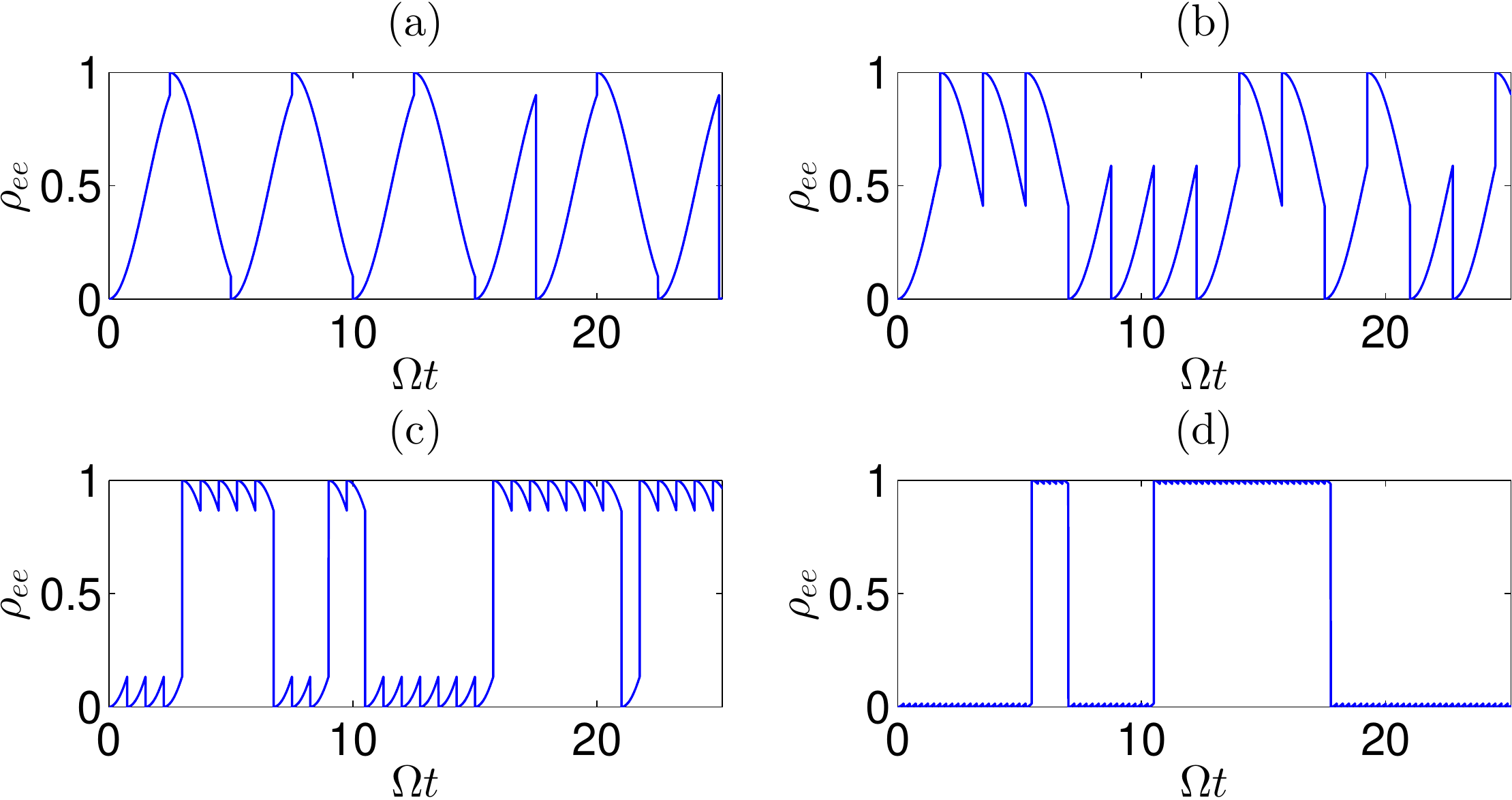}
\caption{\textsl{(Color online) Illustration showing the excited state population as a function of time for a two-level system driven on resonance, and randomly projected at regular intervals according to measurements in the eigenstate basis. The measurement intervals are different in the different panels, (a): $\tau=2.5\Omega^{-1}$, (b): $\tau=1.75\Omega^{-1}$, (c): $\tau=0.75\Omega^{-1}$ and (d): $\tau=0.25\Omega^{-1}$, and show the Quantum Zeno suppression of the coherent Rabi oscillations.
%Realistically one should perform the first measurement at a random point during the first cycle, which is already incorporated in the program. Made with IllustrateZenoDynamics.m.
}}
\label{fig:ZenoEffect}
\end{figure}

As our main example, we consider the estimation of the Rabi frequency $\Omega$ of a driven two-level system subject to projective eigenstate measurements. A similar setup is considered in ref. \cite{Ferrie2013}. Looking at \fref{fig:ZenoEffect}, it is intuitively clear that the Rabi oscillation dynamics is prevented as the interval $\tau$ between measurements decreases, and if we measure too often, the signal holds no information about $\Omega$. If, on the other hand, we measures too rarely we may obtain very little data in any finite probing time $T$. So, what is the optimal value of $\tau$ for the purpose of parameter estimation?
Our expression for the Fisher information provides the $\tau$ maximizing the information in any record obtained in a finite time, and by employing a Bayesian inference protocol \cite{PhysRevA.87.032115} to the measurement outcomes we discuss further the convergence of parameter estimation based on such optimal probing intervals.

The article is organized as follows.
In Sec. \ref{sec:ParFish}, we present a statistical analysis of the Fisher information based on records of projective measurements performed at equidistant times.
In Sec. \ref{sec:ZENO}, we give a simple argument for the QZE in open systems.
In Sec. \ref{sec:twoLevelZeno}, the QZE and Fisher information are discussed in the case of two-level systems.
Sec. \ref{sec:examples} exemplifies the theory in terms of a two-level model with and without inherent dephasing.
In
Sec. \ref{sec:BAYES}, a Bayesian parameter estimation protocol is applied and strategies that reach the asymptotic resolution offered by the Fisher information analysis are investigated. In Sec. \ref{sec:DISCUSS}, we conclude and discuss the relevance of the results for different probing schemes.

\section{Projective measurements and Fisher information}\label{sec:ParFish}
The unconditional evolution of an open quantum system is  in the Born-Markov approximation given by a master equation
$\partial_t\rho(t)=\mathcal{L}\rho(t)$, with a Liouvillian superoperator on Lindblad form,
\begin{align}\label{eq:L}
\mathcal{L}[\rho] = -i[\hat{H},\rho]+\sum_k\left(\hat{c}_k\rho\hat{c}_k^\dagger-\frac{1}{2}\{\hat{c}_k^\dagger\hat{c}_k,\rho\}\right),
\end{align}
where $\hbar=1$, and $\hat{c}_k$ represent different relaxation processes. For time independent operators the solution is formally given by exponentiation, \begin{align}\label{eq:master}
\rho(t)=\e{\mathcal{L}t}[\rho(0)].
\end{align}

If one monitors the time evolution by frequent (continuous) observation the back action conditions the system evolution on the measurement outcomes and the evolution becomes stochastic.
%In the limit of strong and frequent measurements the stochastic evolution dominates, and the dynamics of \eqref{eq:master} is slowed down or frozen out. This this the QZE.

A realistic approach to continuous or frequent measurements should take the finite bandwidth and noise properties of the measurement device into acount, see, e.g.,  \cite{Milburn:88}, but we shall restrict our attention to the ideal case of instantaneous, accurate measurements, repeated at regular intervals $\tau$. The continuous regime approximated by the limit $\tau\rightarrow 0$ thus assumes a corresponding high bandwidth of detection.

While measurement theory can be formulated more generally, for simplicity we restrict our analysis to projective measurements. Each measurement outcome is thus an eigenvalue $\lambda$ of a given operator $\hat{\Lambda}$, occurring with the probability
\begin{align} \label{eq:prob}
P(\lambda)=\mathrm{Tr}(\hat{\Pi}_\lambda\rho(t))\equiv\rho_{\lambda\lambda}(t),
\end{align}
where $\hat{\Pi}_\lambda=\ket{\lambda}\bra{\lambda}$ is a projector on the corresponding eigenstate.
The projection postulate states that the conditional state after a measurement performed on the state $\rho(t)$ at time $t$ is \cite{QMC},
\begin{align}\label{eq:proj}
\rho_\lambda(t) = \frac{\hat{\Pi}_\lambda\rho(t)\hat{\Pi_\lambda}}{P(\lambda)}.
\end{align}
%where the $\hat{\Pi}_\lambda$ are projectors on the subspace of eigenstates with eigenvalue $\lambda$.

\subsection{Fisher information of projective measurement records}
The acquisition of information about unknown physical parameters from a measurement record follows Bayes’ rule: Given the stochastic measurement outcome $D$, the probability that an unknown parameter has a given value $\theta$, is given by the corresponding outcome probability conditioned on the value $\theta$ and its unconditional (prior)
probabilities,
\begin{align}\label{eq:Bayes}
P(\theta|D) = \frac{P(D|\theta)P(\theta)}{P(D)}.
\end{align}
The Cramér Rao bound (CRB) \cite{Cramer},
\begin{align}\label{eq:CRB}
[\Delta S(\theta)]^2\geq \frac{1}{F(\theta_0)},
\end{align}
yields a lower bound of the statistical variance $[\Delta S(\theta)]^2$ on any unbiased estimate of the unknown quantity $\theta$ (with the true value $\theta_0$).
Here,
\begin{align}
\label{eq:fisher}
F(\theta) &=
%-\sum_D P(D|\theta)\partial_{\theta}^2 \ln P(D|\theta)
%\\
%&=
 -\mathbb{E}\left[\partial_{\theta}^2 \ln P(D|\theta)\right],
\end{align}
is the Fisher information which quantifies the dependence of the data outcome probabilities on the unknown parameter. The Fisher information and the Cramér Rao bound yield the asymptotic precision of the best possible estimate, and normally assumes a large number of repetitions $N$ of the experiment (and a corresponding factor $1/N$ in \eqref{eq:CRB}). In our analysis a self-averaging occurs due to  the repetitive measurements, and we shall find the corresponding dependence of $F(\theta)$ on $N$ and, equivalently, on the total time of the probing $T$.

%\subsection{Fisher information of projective measurement records}
A data record of $N =T/\tau$ projective measurement outcomes obtained at times $t_j=j\tau$, $D=\{\lambda_j\}_{j=1}^N$, occurs with the probability
\begin{align}
&P(D|\theta) =
\prod_j \Tr{\ket{\lambda_j}\bra{\lambda_j}\e{\mathcal{L}\tau} [\rho(t_{j-1})]}
\nonumber
\\
%&=
%\prod_j \frac{\Tr{\ket{\lambda_j}\bra{\lambda_j}\e{\mathcal{L}\tau} [\ket{\lambda_{j-1}}\bra{\lambda_{j-1}}\rho(t_{j-2}+\tau)\ket{\lambda_{j-1}}\bra{\lambda_{j-1}}]}}{\Tr{\ket{\lambda_{j-1}}\bra{\lambda_{j-1}}\rho(t_{j-2}+\tau)}}
%\\
%\nonumber
&=
\prod_j \Tr{\ket{\lambda_j}\bra{\lambda_j}\e{\mathcal{L}\tau} [\ket{\lambda_{j-1}}\bra{\lambda_{j-1}}]},
\end{align}
assuming that the initial state is one of the eigenstates, $\rho(0)=|\lambda_0\rangle\langle \lambda_0|$.
Since by \eqref{eq:proj} the system is always projected (reset) in one of the eigenstates upon detection, the probability for an entire data record factors into a product of conditional probabilities $P(\lambda_{j}|\lambda_{j-1},\theta)=\Tr{\ket{\lambda_j}\bra{\lambda_j}\e{\mathcal{L}\tau} [\ket{\lambda_{j-1}}\bra{\lambda_{j-1}}]}$, where we recall that the time evolution operator (\ref{eq:master}) depends on $\theta$. The data record is fully represented by the set of numbers $n_{lm}$ counting the occurrences of subsequent detection in states $(\lambda_{j-1},\lambda_{j})=(\lambda_m,\lambda_{l})$.
The mean number of such events during a total of $N$ measurements is
$\mathbb{E}[n_{lm}]=NP(\lambda_{m}|\lambda_l,\theta)P(\lambda_l|\theta)$, where the probability that the average measurement will yield $\lambda_l$ may be calculated as
$P(\lambda_l|\theta)= \rho_{\lambda_i\lambda_i}^{\mathrm{st}}$ with $\rho^{st}$ the stationary (steady state) solution of the non-selective evolution in \eqref{eq:master}, where the average measurement back action is included as a dephasing of the atomic coherence at a rate $\propto \tau^{-1}$ (See Sec. \ref{sec:dep} below).

The conditional probability for subsequent measurements, $P(\lambda_{j}=\lambda_{m}|\lambda_{j-1}=\lambda_l,\theta)$ does not depend on $j$, and
with a total number of measurements $N=\sum_{lm}n_{lm}$, the probability in Eqs. (\ref{eq:Bayes},\ref{eq:fisher}) for the data record $D=\{n_{lm}\}$ is a multinomial distribution,

\begin{align}\label{eq:probabilities}
P(D|\theta) = \prod_{lm}P(\lambda_{m}|\lambda_l,\theta)^{n_{lm}}.
%\frac{N }{\prod_{lm}n_{lm}!}\prod_{lm}.
\end{align}
Utilizing $\sum_m P(\lambda_{m}|\lambda_l,\theta)=1$,
we readily obtain the Fisher information for a sequence of $N$ projective measurements, which we can represent as
\begin{align}\label{eq:FisherGeneral}
\frac{F_\tau(\theta)}{N} = \sum_{lm} \frac{(\partial_\theta P(\lambda_{m}|\lambda_l,\theta))^2}{P(\lambda_{m}|\lambda_l,\theta)}P(\lambda_l|\theta).
\end{align}
The result \eqref{eq:FisherGeneral} applies to any parameter estimate from a record of projective measurements performed at constant intervals $\tau$.

The quantity of relevance in this article is the Fisher information obtained for a given, long, interrogation time, $T=N\tau$,
and to address the asymptotic precision as well as the optimal value of $\tau$, we shall consider the Fisher information per time $F_\tau(\theta)/T=F_\tau(\theta)/N\tau$.

%The conditional probabilities follow from \eqref{eq:prob},
%$P(\lambda_j|\lambda_{i},\theta)=\left.\rho_{\lambda_j\lambda_j}(\tau)\right|_{\rho(0)=\ket{\lambda_i}\bra{\lambda_i}}$, 
%The probabilities $P(\lambda_i|\theta)

\section{Zeno inhibited evolution}\label{sec:ZENO}
Following Peres \cite{Peres}, we give now a simple account of the Quantum Zeno effect.
Let $\rho_0$ denote a pure state of a quantum system subject to evolution by the master equation \eqref{eq:master}. The probability that the system will be found in its initial state after a short time $\tau$ is
\begin{align}\label{eq:P0}
P_0 \simeq 1+a\tau+b\tau^2,
\end{align}
where $a = \Tr{\rho_0\mathcal{L}[\rho_0]}$ and $b = \frac{1}{2}\Tr{\rho_0\mathcal{L}^2[\rho_0]}$.
In cases where $a\neq 0$, the dominating contribution to $P_0$ is linear in $\tau$, and the survival probability in the initial state after $N$ time steps and projective measurements is
\begin{align}\label{eq:linear}
P^{\mathrm{linear}}_0(N) \simeq\left[1+a\frac{T}{N}\right]^N,
\end{align}
yielding an exponential decay law $P_0^{\mathrm{linear}}(N\rightarrow \infty)=\exp\left( aT\right)$ for large $N$.
% In this case, no QZE is observed.

In a number of cases, $a=0$ (See Sec. \ref{sec:twoLevelZeno} below). Then, if the measurement is performed $N$ times at intervals $\tau=T/N$, the probability that all measurements yield the initial state reads
\begin{align}\label{eq:ZenoP0}
P_0^{\mathrm{quadratic}}(N) \simeq \left[1+b\left(\frac{T}{N}\right)^2\right]^N,
\end{align}
In the limit $N\rightarrow \infty$, $P_0^{\mathrm{quadratic}}(N)$ tends to $1$: The QZE freezes the system dynamics.

At small but finite $\tau$, it is instructive to set $(T/N)^2=\tau (T/N)$ in \eqref{eq:ZenoP0} and keep $\tau$ constant. Then, taking the limit $P_0^{\mathrm{quadratic}}(N\rightarrow \infty) = \exp\left(\tau b T\right)$. The probability to find the system in its initial state thus decays on the Zeno time scale $\tau_Z$, given by  $\tau_Z^{-1}\equiv-\tau b$. For closed system dynamics governed by a Hamiltonian $\hat{H}$,
$a$ vanishes, but we obtain $\tau_Z^{-1}=\tau (\braket{\hat{H}^2}- \braket{\hat{H}}^2)$ \cite{Peres}.

\section{QZE and Fisher information for a probed two-level system}\label{sec:twoLevelZeno}
We shall now investigate how the Zeno effect affects precision probing by considering a two-level system driven by a laser field.
In a frame rotating with the frequency of the laser field, the Hamiltonian reads
\begin{align}\label{eq:H1}
\hat{H} =
-\delta \ket{e}\bra{e} +\frac{\Omega}{2}\left(\ket{e}\bra{g}+\ket{g}\bra{e}\right),
\end{align}
where $\delta$ is the atom-laser detuning and $\Omega$ is the laser-Rabi frequency.

Following the previous discussion, projective measurements in the atomic eigenstate basis, $\hat{\Pi}_{g}=\ket{g}\bra{g}$ and $\hat{\Pi}_{e}=\ket{e}\bra{e}$, are performed at intervals $\tau$.
The conditional, stochastic evolution is exemplified in \fref{fig:ZenoEffect} for different durations of the interval $\tau$ between measurement during which the evolution given by (\ref{eq:H1}) is unitary. Decreasing $\tau$ gradually inhibits the coherent dynamics, and eventually leads to quantum jumps occurring at the rate $\tau_Z^{-1} = \tau\Omega^2/4$.

To determine the Fisher information in \eqref{eq:FisherGeneral},
%\subsection{Population measurements in a two-level system}
%Specializing to population measurements in a two-level system as introduced in Sec. (\ref{sec:twoLevelZeno}),
we note that the $\lambda_j$ take only two values corresponding to measuring the ground ($\lambda_j=g$) or excited ($\lambda_j=e$) state respectively.
In this case,
\begin{align}\label{eq:F2}
\frac{F_\tau(\theta)}{N} &= \frac{P(g|\theta)(\partial_\theta P(g|g,\theta))^2}{P(g|g,\theta)(1-P(g|g,\theta))}
\nonumber
\\
&\quad\quad+
\frac{P(e|\theta)(\partial_\theta P(e|e,\theta))^2}{P(e|e,\theta) (1-P(e|e,\theta))},
\end{align}
where $P(g|\theta)$ and $P(e|\theta)$ are the populations of the ground and excited state, in the unconditional steady state of the system. Absent energy relaxing processes these are both $\frac{1}{2}$.

Due to the symmetry of the problem, the only relevant information in the measurement record concerns how often consecutive measurements yield identical or different results, and we have
\begin{align}\label{eq:FBino}
\frac{F_\tau(\theta)}{N} = \frac{(\partial_\theta P(g|g,\theta))^2}{ P(g|g,\theta)(1-P(g|g,\theta))},
\end{align}
which can be recognized as the Fisher information for binomially distributed data.
\\\\
%\begin{align}
%\frac{F_\tau(\theta)}{N} \simeq  \left(\partial_\theta 1+\Tr{\rho_0\mathcal{L}[\rho_0]}\tau+\frac{\Tr{\rho_0\mathcal{L}^2[\rho_0]}}{2}\tau^2\right)^2
%\end{align}
%\begin{align}
%\frac{F_\tau(\theta)}{N} \simeq  \frac{(\partial_\theta \Tr{\rho_0\mathcal{L}[\rho_0]})^2\tau^2+2\partial_\theta \Tr{\rho_0\mathcal{L}[\rho_0]}\tau^3+\left(\partial_\theta \frac{1}{2}\Tr{\rho_0\mathcal{L}^2[\rho_0]}\tau^2\right)}
%{\Tr{\rho_0\mathcal{L}[\rho_0]}\tau +\frac{1}{2}\Tr{\rho_0\mathcal{L}^2[\rho_0]}\tau^2},
%\end{align}

For very short durations between measurements $P(g|g,\theta)$ is on the form of \eqref{eq:P0} with $\rho_0=\ket{g}\bra{g}$, and from \eqref{eq:FBino} the Fisher information is
\begin{align}\label{eq:17}
\frac{F_\tau(\theta)}{N} \simeq \frac{(\partial_\theta a)^2\tau^2 +2(\partial_\theta a)(\partial_\theta b)\tau^3+(\partial_\theta b)^2\tau^4}{a\tau+(a^2+b)\tau^2},
\end{align}
If $a=0$ the QZE freezes out the system dynamics in the limit of $\tau \rightarrow 0$, and
$F_\tau(\theta)/T \simeq \tau(\partial_\theta b)^2/b\rightarrow 0$, so despite the growing number of measurements, $N=T/\tau\rightarrow \infty$ the data holds vanishing information on $\theta$.
If, on the other hand, $a\neq0$, we find a constant Fisher information per time in the limit of $\tau \rightarrow 0$, $F_\tau(\theta)/T \simeq (\partial_\theta a)^2/a$.
Note, that if $a$ is independent of $\theta$ the Fisher information may still vanish even though the dynamics is not frozen out for $\tau \rightarrow 0$.

For simplicity, we restricted this discussion to the case of a two-level system, but the relation between a vanishing linear term in the survival probability
\eqref{eq:P0}
and Quantum Zeno inhibited parameter estimation equally applies to the general case of
\eqref{eq:FisherGeneral}.

% for fixed $N$.
%The challenge in this calculation was to realize that the record just samples a binomial distribution with the given probabilities.

%and the data record consists of independent results.
%We assign probabilities $P(\lambda_{i+1}|\lambda_i)$ to these combined events, and let $M$ denote the number of events where $\lambda_i=\lambda_{i+1}$. Then from the independence
%\begin{align}
%P(D|\theta) &= \prod_i^N P(\lambda_{i+1}|\lambda_i,\theta)
%\\
%&=  P(\lambda_{i+1}=\lambda_i|\lambda_i)^M P(\lambda_{i+1}\neq %\lambda_i|\lambda_i)^{N-M},
%\end{align}
%which is just a binomial distribution (without the coefficient ${{N}\choose{M}}$?)!
%From the symmetry we may furthermore assign all probabilites by letting $\lambda_i=-1$, so that $P(\lambda_{i+1}=\lambda_i|\lambda_i) = \left.\rho^\theta_{gg}(\tau)\right|_{\rho^\theta(0)=\ket{g}\bra{g}} \equiv p$ and $P(\lambda_{i+1}=\lambda_i|\lambda_i)=1-p$. Here $\rho^\theta(t)$ is the no probe state evolved with the given value of $\theta$.
%Then $\mathbb{E}[M] = N p$, and remembering that $M$ stems from an actual record and thus has no explicit $\theta$-dependence (like the $n_k$ is photon counting case) we find from \eqref{eq:fisher} the Fisher information

\section{Examples}\label{sec:examples}
In this section we exemplify the Zeno inhibited Fisher information for
Rabi-frequency estimation in a driven two-level system. We first discuss the case of unitary dynamics between measurements, and we then turn to a model including dephasing.

\subsection{Driven system with no dephasing}\label{sec:noDephasing}
For the two-level system with Hamiltonian \eqref{eq:H1} the time evolution may be solved analytically,
\begin{align}
%\left.\rho_{gg}(\tau)\right|_{\rho(0)\ket{g}\bra{g}}
P(g|g,\Omega)= \frac{1}{2}\left[1+\frac{\delta^2+\Omega^2\cos \chi \tau}{\chi^2} \right]
\end{align}
where $\chi = (\Omega^2+\delta^2)^{1/2}$ is the generalized Rabi-frequency.
By \eqref{eq:FBino} the corresponding Fisher information per measurement for Rabi-frequency estimation is
%\footnote{AnalyticFisher.nb}
\begin{align} \label{fnod}
\frac{F_\tau(\Omega)}{N} = \frac{\left(\Omega^2\chi\tau\cos\frac{\chi\tau}{2}
+
2\delta^2\sin\frac{\chi\tau}{2}\right)^2}{
\chi^4 (\delta^2 + \Omega^2\cos^2\frac{\chi\tau}{2})}.
\end{align}

\begin{figure}
\centering
\includegraphics[trim=0 0 0 0,width=0.9\columnwidth]{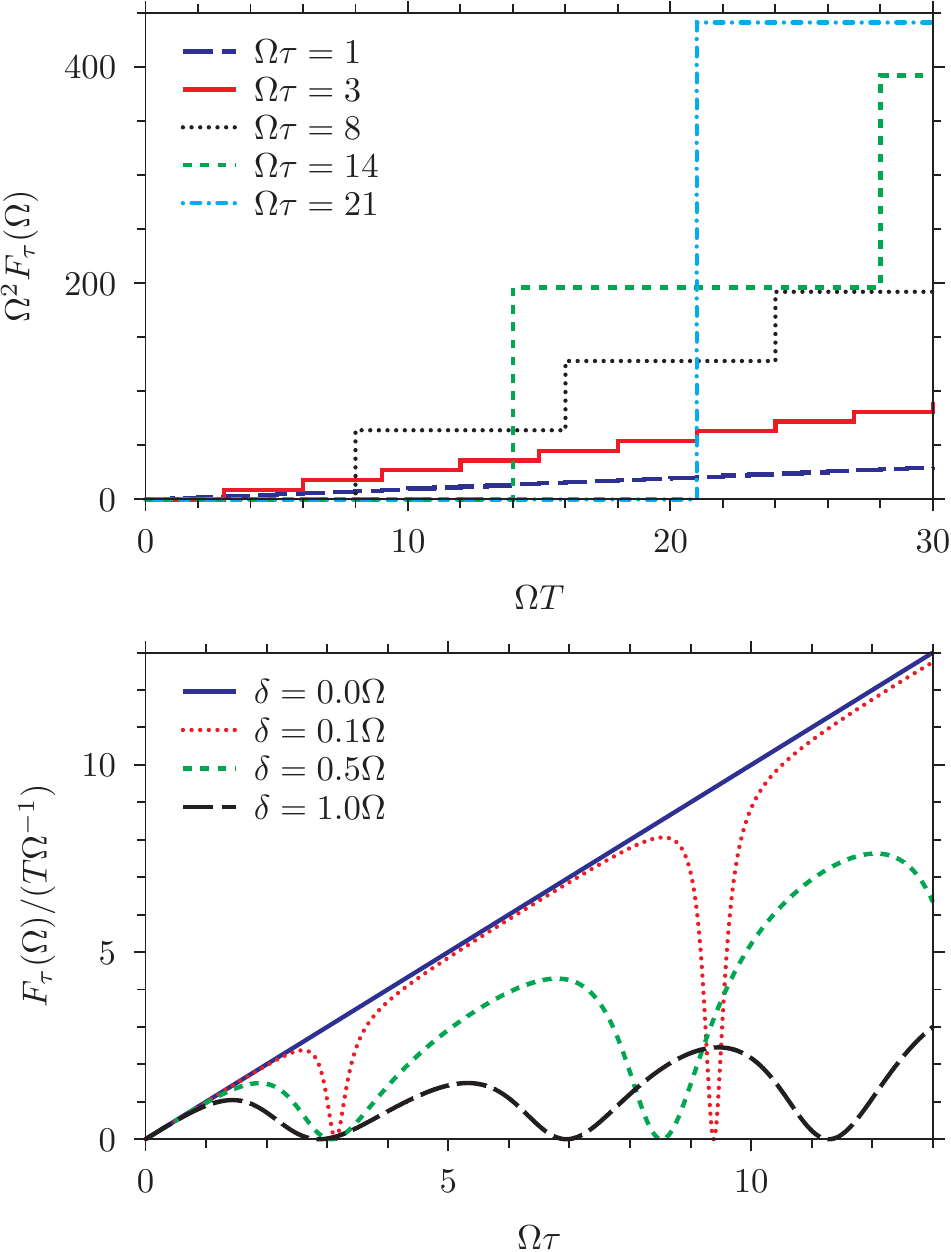}
\caption{\textsl{(Color online)
 Upper panel: The Fisher information $F_\tau(\Omega)=\tau T$ for Rabi frequency estimation in a resonantly driven two-level system as a function of the total probing time. Results are shown for different durations of the intervals between measurements.
Lower panel: The Fisher information \eqref{fnod} per time for Rabi frequency estimation as a function of the duration of the intervals between measurements  assuming $\tau \ll T$. Results are shown for different detunings.
%Might put in $\tau_Z$ Made with FisherRabiAnalytical.m
}}
\label{fig:FisherAnalytic}
\end{figure}
In the upper panel of \fref{fig:FisherAnalytic}, we show how the Fisher information grows as a function of the total interrogation time $T = N\tau$ for different values of $\tau$. Information is only obtained when measurement outcomes are acquired, and the Fisher information therefore grows in steps.
While the frequency of the steps increases when the durations between measurements are decreased, the figure clearly shows a reduction of both the height of the individual steps and the effective slope of the Fisher information as function of  time $T$. In the lower panel of \fref{fig:FisherAnalytic}, the slope $F_{\tau}/T$ is shown as function of $\tau$ for different values of the laser detuning and $\tau\ll T$ .
On resonance ($\delta=0$) \eqref{fnod} reduces to the Fisher information per measurement,  $F_\tau(\Omega)/N = \tau^2$, independent of the actual value of $\Omega$ (equivalent to a linear relationship between $F_\tau(\Omega)/(T\Omega^{-1})$ and $\Omega\tau$). For a fixed total time, the Rabi oscillation dynamics is frozen out and as anticipated by the discussion following \eqref{eq:17} any $\Omega$-estimate is inhibited by the QZE in the limit $\tau\rightarrow 0$.

\begin{figure}
\centering
\includegraphics[trim=0 0 0 0,width=0.9\columnwidth]{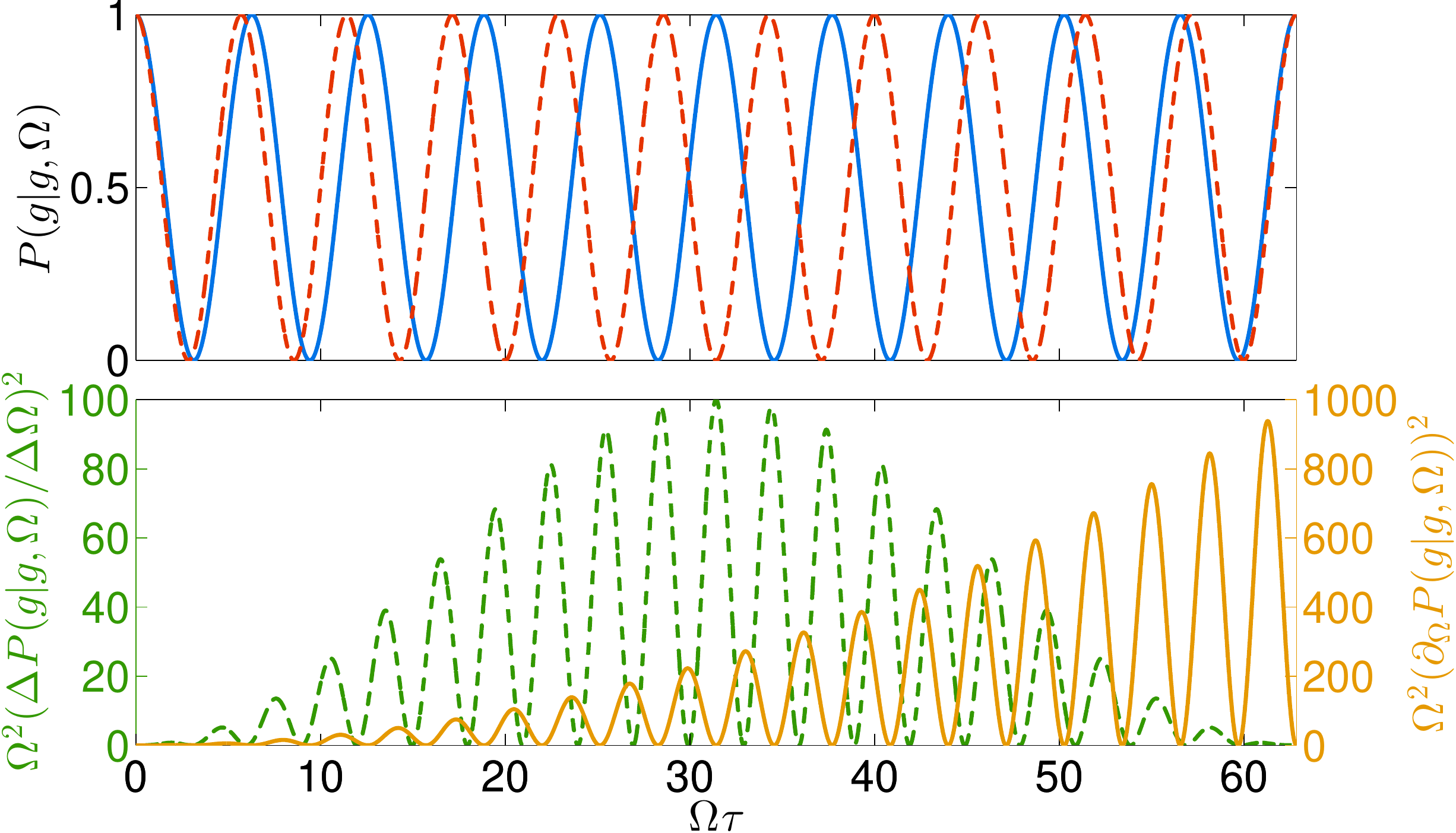}
\caption{\textsl{(Color online) Upper panel: The probability of finding a two-level system driven on resonance in the ground state as a function of time for $\Omega_1=\Omega$ (blue curve) and $\Omega_2=1.1\Omega$ (dashed, red curve).
Lower panel: The dashed, green curve shows the squared difference quotient, $\left[P(g|g,\Omega_2)-P(g|g,\Omega_1)\right]/\left[\Omega_2-\Omega_1\right]$, between the probabilities in the upper panel while the orange curve shows the squared  differential quotient, $\partial_\Omega P(g|g,\Omega)$, corresponding to the limit of infinitesimally close Rabi frequencies.
%Made with FisherRabiNumericalDiff.m
}}
\label{fig:finiteDelta}
\end{figure}

To understand the behaviour for large $\tau$, we consider the numerator in \eqref{eq:FBino}.
In the upper panel of \fref{fig:finiteDelta} $P(g|g,\Omega)$ is shown for two slightly different values of $\Omega$.
The two candidate values lead to solutions evolving at different frequencies acquiring the same phase after a certain time. This effect contributes the envelope to the difference quotient seen in the lower plot (dashed, orange curve). The smaller the difference between the Rabi frequencies, the later the peak appears in the envelope function, and in the limit of infinitesimally close Rabi frequencies, their Rabi oscillations will only come back in phase after an infinitely long time. As the Fisher information relates to infinitesimally different hypotheses, a single measurement at time $\tau=T$ provides the optimal distinction, and the limiting procedure of preparation of an initial state followed by a single read out at the final time is favoured. Indeed, the quadratic dependence on $\tau$, causing the Zeno suppression for short measurement intervals, is responsible for the advantageous Heisenberg scaling when $\tau=T$ \cite{PhysRevLett.114.210801}.

%Conversely, for large $\tau$, the Fisher information per measurement $N$ grows quadratically with $\tau$ (Heisenberg scaling), favouring the limiting procedure of preparation of an initial state followed by a single read out at $\tau=T$.

%Notice that the differential is zero when the state $\rho(\tau)$ is at the pole of the Bloch sphere, while it has local maxima at the equator. On resonance this is countered by an oscillating denominator in \eqref{eq:FBino} vanishing at the poles, and resulting in a non-oscillatory Fisher information.

\subsection{Driven system with dephasing}\label{sec:dep}

Real physical systems experience dephasing due to e.g. magnetic noise, off-resonant light scattering and collisions with back ground gas \cite{BoseImplementation,PhysRevA.41.2295}.
To study how such effects affect parameter estimation, we introduce a transversal dephasing of the atomic coherence at a rate $\gamma$ with the Lindblad operator $\sqrt{\gamma}\hat{\sigma}_z$ in \eqref{eq:L}.
The ground and excited states are eigenstates of $\hat{\sigma}_z$, and $\Tr{\rho_0\mathcal{L}[\rho_0]}$ still vanishes so that by \eqref{eq:ZenoP0} the QZE is present in the system.

The system evolution may be solved analytically, and on resonance one finds
\begin{equation}\label{eq:rhogg}
 P(g|g,\Omega)=\left\{
\begin{array}{ll}
      \frac{1}{2}+\frac{\mathrm{e}^{-\gamma \tau}}{2}\left(\frac{\gamma}{\omega}\sin \omega \tau +\cos \omega \tau \right) & \Omega \neq \gamma\\
      \frac{1}{2}+\frac{\mathrm{e}^{-\Omega \tau}}{2}\left(\Omega \tau+1 \right) & \Omega = \gamma,
\end{array}
\right.
\end{equation}
where $\omega=\sqrt{\Omega^2-\gamma^2}$.

\begin{figure}
\centering
\includegraphics[trim=0 0 0 0,width=1\columnwidth]{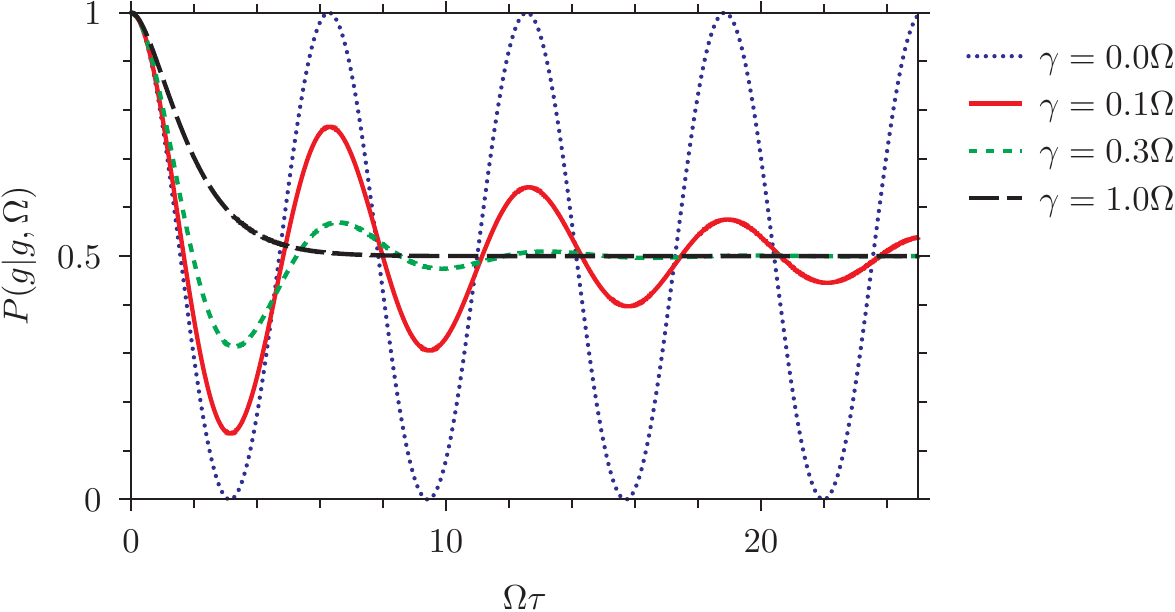}
\caption{\textsl{(Color online) The time evolution of $P(g|g,\Omega)$ for a two-level system driven on resonance. Results are show for values of $\gamma$ ranging from no dephasing ($\gamma=0$) to critical damping of the Rabi oscillations ($\gamma=\Omega$).
%Made with NoProbeEvolution.m. Put only one value.
}}
\label{fig:NoProbeEvolution}
\end{figure}
Results are shown in \fref{fig:NoProbeEvolution} which depicts the time evolution of $P(g|g,\Omega)$ for different values of $\gamma$. Due to the dephasing, the oscillations are damped over time, and it is intuitively clear that it is no longer optimal to wait as long as possible between measurements.

Using \eqref{eq:FBino} we determine the Fisher information per measurement
\begin{align}\label{eq:Fisher_dephasing}
\frac{F_\tau(\Omega)}{N} = \frac{-2 \Omega^2\left(\gamma \omega \tau \cos \omega\tau -
        (\gamma +\omega^2 \tau) \sin\omega\tau \right)^2}
       {\omega^4 (\Omega^2 -
       2 \omega^2\mathrm{e}^{2 \gamma \tau}  + (\omega^2- \gamma^2) \cos 2\omega\tau
          +
       2 \gamma\omega \sin 2 \omega \tau)},
\end{align}
valid for $\Omega \neq \gamma$ as well as in the limit $\Omega \rightarrow \gamma$.
%For $\Omega = \gamma$ the differential is not well-defined, and one must be a little more careful, when calculating the Fisher information. It is found that the analytical and numerical results does not quite match if \eqref{eq:critical} is used.

\begin{figure}
\includegraphics[trim=0 0 0 0,width=0.8\columnwidth]{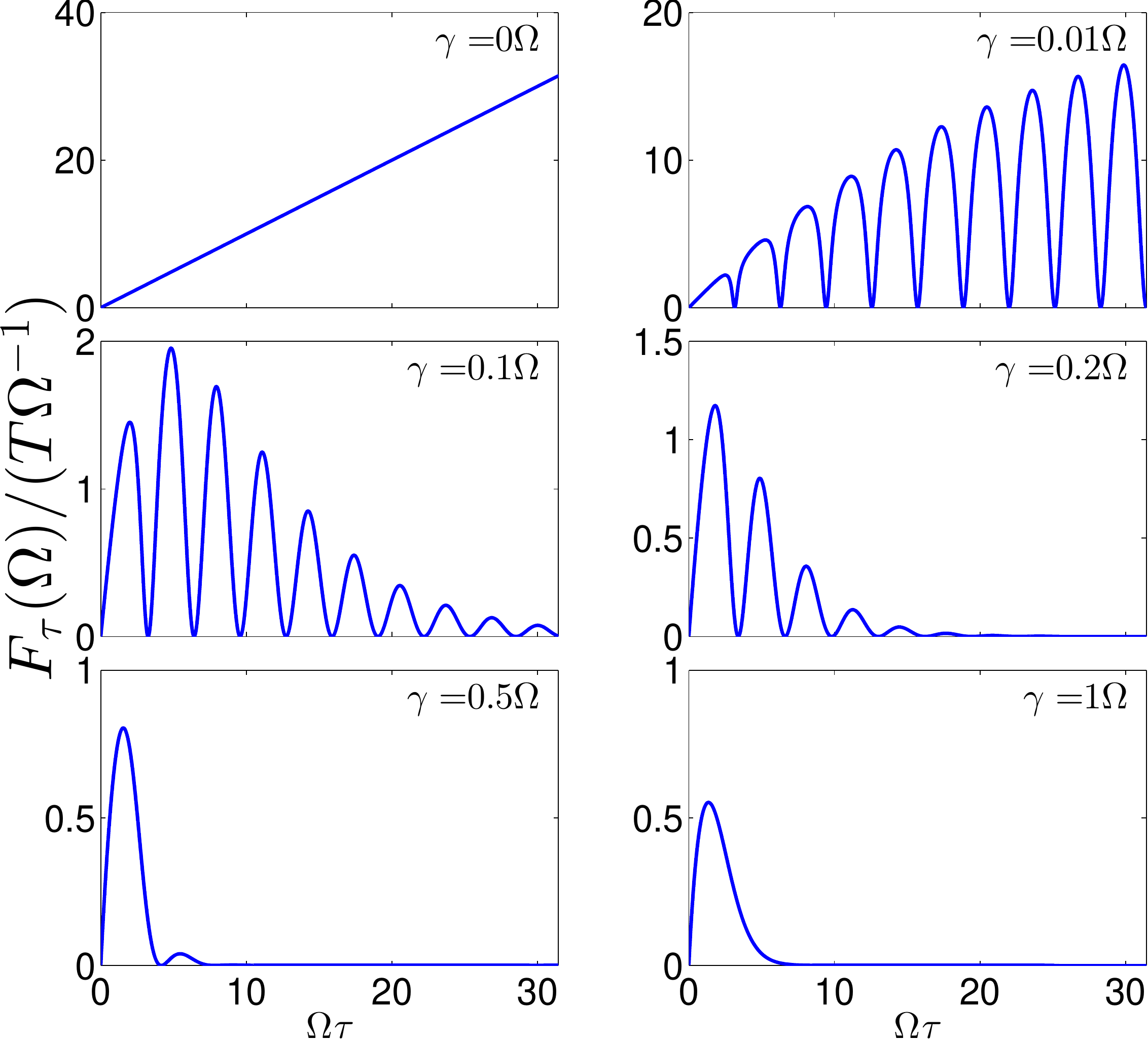}
\includegraphics[trim=0 0 0 0,width=0.9\columnwidth]{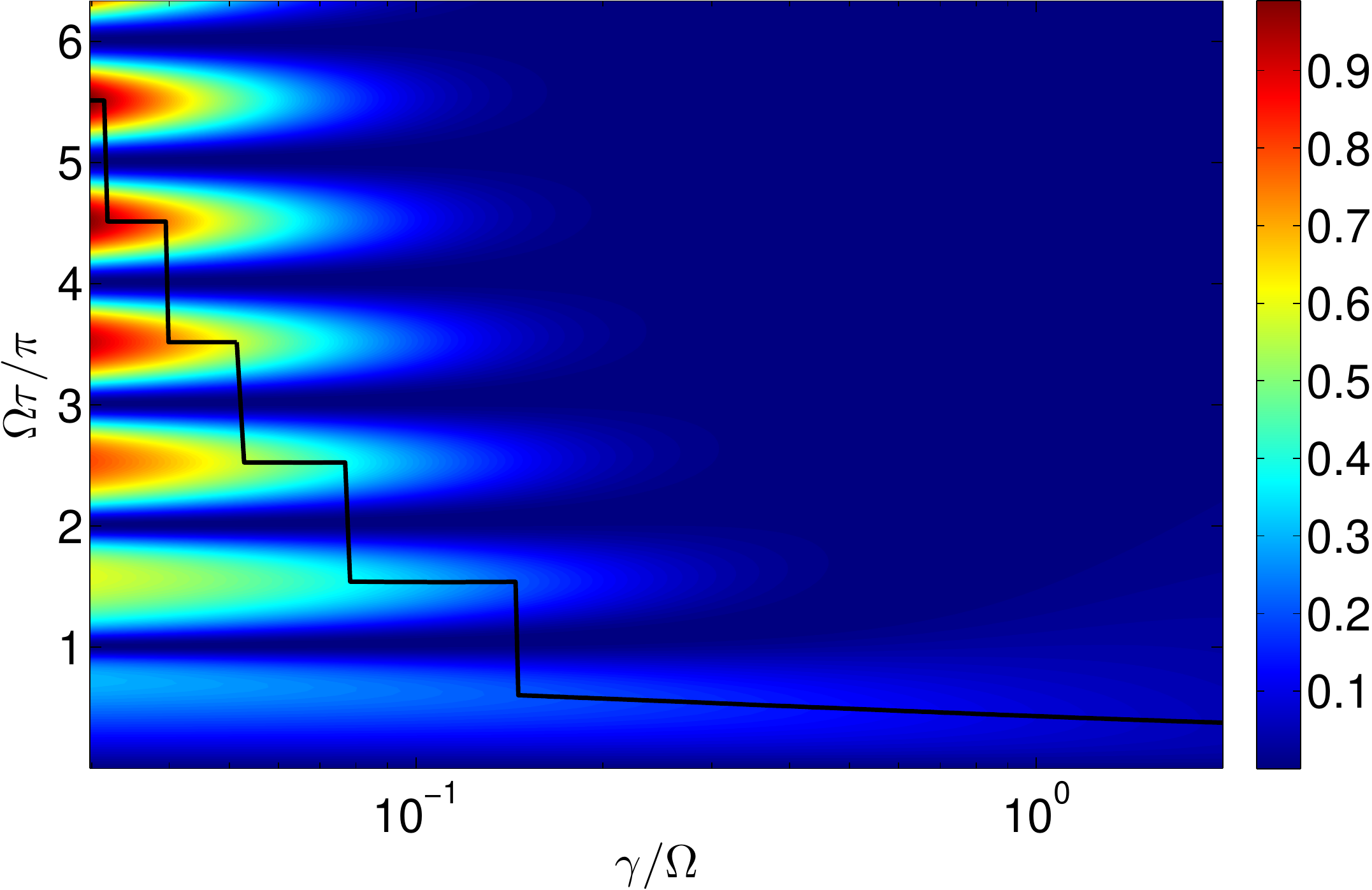}
\caption{\textsl{(Color online)
The Fisher  information \eqref{eq:FBino} per time for Rabi frequency estimation in a two-level system driven on resonance. In the six upper panels, results are shown  as a function of the duration of the intervals between measurements and values of $\gamma$ ranging from no dephasing ($\gamma=0$) to critical damping of the Rabi oscillations ($\gamma=\Omega$).
The color plot pictures the Fisher information as a function of both $\tau$ and $\gamma$. The color scale is in units relative to the maximum value. The black curve in the plot tracks the value of $\tau$ maximizing the Fisher information for any given value of $\gamma$. For weak damping, the maxima occur close to $\omega\tau=\pi(n+\frac{1}{2})$, where $n$ is an integer.
%Made with FisherRabiNumerical.m
%Notice the close resemblance to Mott-Superfluid plot in Hubbard model. Made with OptimalKappa.m. Remove two upper plots. (Change axis index later to make more transparent)
}}
\label{fig:FisherDephasing}
\end{figure}

%It is seen how the dephasing is moving the maximum in the differential to finite $\tau$ while killing the oscillations in the denominator.
In the six upper panels of \fref{fig:FisherDephasing}, the Fisher information per time for different dephasing rates is depicted. Note the oscillatory behaviour, and that it is indeed peaked for finite measurement intervals.

From \eqref{eq:Fisher_dephasing} one finds, in the strong driving regime,
\begin{align}
\frac{F_\tau(\theta)}{T} = \tau\mathrm{e}^{-2\gamma \tau}\frac{\sin^2 \omega\tau}{1-\mathrm{e}^{-2\gamma \tau}\cos^2 \omega \tau}
\quad \quad \text{for } \Omega \gg \gamma.
\end{align}
For small $\tau$ the envelope $\tau\mathrm{e}^{-2\gamma \tau}$ yields the same linear dependence as in the undamped case caused by the Quantum Zeno effect, while for larger $\tau$ dephasing suppresses the information.
%, reflecting a trade off between information loss by inherent dephasing and the QZE.
%Notice how the carrier is more complicated in the linear regime, while it is dominated by $\sin^2 \omega$ for higher $\tau$.
\\
%$\clubsuit$ KLAUS: Jeg synes ikke man faar saa meget ud af dette afsnit og denne figur.
%To investigate the Zeno-regime further we show in \fref{fig:tauDerivative} the derivative of the Fisher information per time with respect to $\tau$ for different dephasing rates. The linearity of $F_\tau(\Omega)$ as $\tau\rightarrow 0$, confirms the presence of the QZE for all values of $\gamma$. Notice, however, that as $\gamma$ increases the range of values of $\tau$, for which the QZE is appreciable, diminishes.
%\begin{figure}
%\centering
%\includegraphics[trim=0 0 0 0,width=1\columnwidth]{dF.pdf}
%\caption{\textsl{(Color online) The derivative with respect to $\tau$ of the Fisher  information \eqref{eq:FBino} per time for Rabi frequency estimation. Results are shown for values of $\gamma$ ranging from no dephasing ($\gamma=0$) to overdamped Rabi oscillations ($\gamma=2\Omega$). Notice, that in all cases the derivative approaches unity when $\tau\rightarrow 0$.
%%Made with AnalyticFisher.nb.
%}}
%\label{fig:tauDerivative}
%\end{figure}
%$\clubsuit$

The lower panel of \fref{fig:FisherDephasing} shows in color the Fisher information as a function of $\gamma$ and $\tau$, and the black curve indicates the values of $\tau$ that maximize the information for any given $\gamma$. For high $\gamma$, phase information is only significantly maintained during the first $\pi/2$ of the rotation and quite frequent measurements are optimal. As $\gamma$ decreases, additional $\pi$ rotations are favourable between measurements, and since the sensitivity to $\Omega$ by population measurements is highest when both populations are close to $\frac{1}{2}$, the optima occur around $\omega\tau=\pi(n+\frac{1}{2})$ where $n$ is an integer. The optimal measurement intervals change in steps as the dephasing rate becomes smaller and smaller.

%It is all the time a competition between envelope and carrier in the differential, and the jumps occur once the envelope maximum moves from one peak to the next.
%\begin{figure}
%\centering
%\includegraphics[trim=0 0 0 0,width=0.8\columnwidth]{/home/alexander/Copy/PhD/QuantumZenoEffect/MatLab/OptimalKappa.pdf}
%\includegraphics[trim=0 0 0 0,width=0.9\columnwidth]{Fgammakappa.pdf}
%\caption{\textsl{(Color online) The color plot shows the Fisher  information \eqref{eq:FBino} per time for Rabi frequency estimation in a two-level system driven on resonance as a function of the probing interval $\tau$ and transversal damping coefficient $\gamma$. The black curve tracks the value of $\tau$ maximizing the Fisher information for any given value of $\gamma$. Note, that the maxima occur close to $\omega\tau=\pi(n+\frac{1}{2})$ where $n$ is an integer.
%Notice the close resemblance to Mott-Superfluid plot in Hubbard model. Made with OptimalKappa.m. Remove two upper plots. (Change axis index later to make more transparent)
%}}
%\label{fig:OptimalKappa}
%\end{figure}
%Notice how this behaviour resembles phase transition behaviour. Once we reach the final peak (first round on the Bloch sphere) we no longer have "phase" transitions in information.

\section{Bayesian inference}\label{sec:BAYES}
In the sections above we investigated the theoretical optimal sensitivity based on projective measurements. We shall now address the achievement of this sensitivity by an explicit estimation strategy.
Given the outcome at the time $t_j$=$j\tau$, Bayes rule
\eqref{eq:Bayes} yields an update for the probability $P(\theta)$ assigned to different values of $\theta$ prior to the detection event. The update rule is applied at each detection and thus acts as a filter multiplying all the probability factors assigned to the outcomes \textit{actually occurring} during the experiment on the prior, say, uniform probabilities assigned to the candidate values for the parameter $\theta$ \cite{PhysRevA.87.032115} .
%Note that $P(D)$ does not depend on $\theta$

\begin{figure}[h]
\centering
\includegraphics[trim=0 0 0 0,width=1.0\columnwidth]{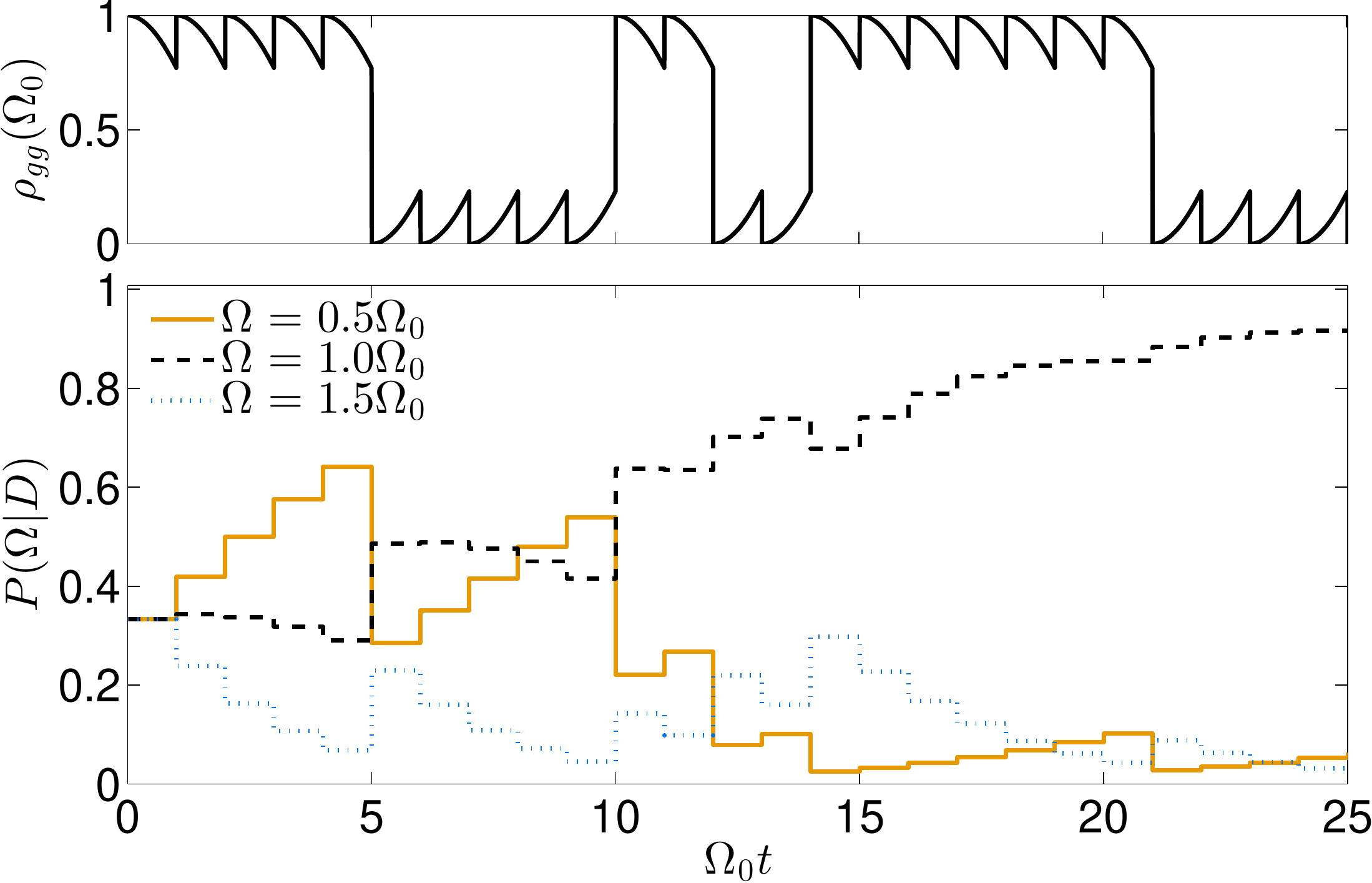}
\caption{\textsl{(Color online) The top panel shows the ground state population resulting from a quantum trajectory simulation with $\Omega=\Omega_0$ and $\gamma=\delta=0$. The lower panel shows the evolution of the probabilities $P(\Omega|D)$ for three candidate values of the Rabi frequency, conditioned on the measurement record in the upper plot. Results are shown for $\Omega_0\tau=1$.
%Made with Bayes.m.
}}
\label{fig:Bayes1}
\end{figure}

We illustrate this principle in the lower panel of \fref{fig:Bayes1} by propagating probabilities based on three candidate values of the Rabi frequency used to generate the trajectory in the upper panel.
The trajectory is dominated by intervals of projections to the same state in consecutive measurements (the QZE), consistent with a vanshing Rabi frequency and hence causing an increase in the probability for the lowest value of $\Omega$.  Due to the finite value of $\tau$, however, the system performs quantum jump like dynamics between the eigenstates at a rate $\sim \tau_Z^{-1}$, favoring the higher values of $\Omega$ in the filtering process.
Eventually, after a long sequence of measurements, the full record is only consistent with the true value of $\Omega$.

\begin{figure*}
\centering
\subfloat[\label{subfig-1:dummy}]{%
      \includegraphics[trim=0 0 0 0,width=0.66\columnwidth]{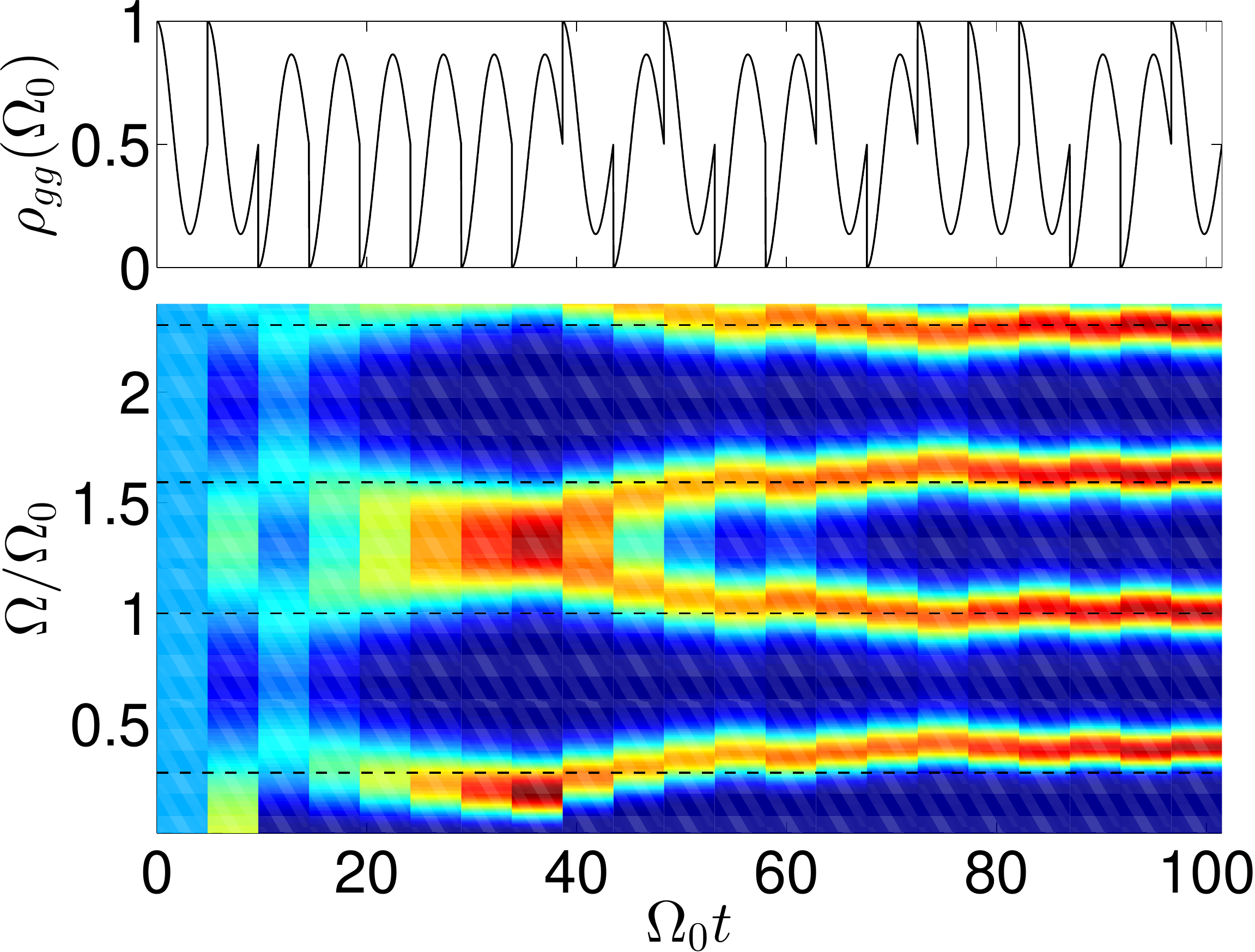}
    }
%\hfill
\subfloat[\label{subfig-2:dummy}]{
\includegraphics[trim=0 0 0 0,width=0.66\columnwidth]{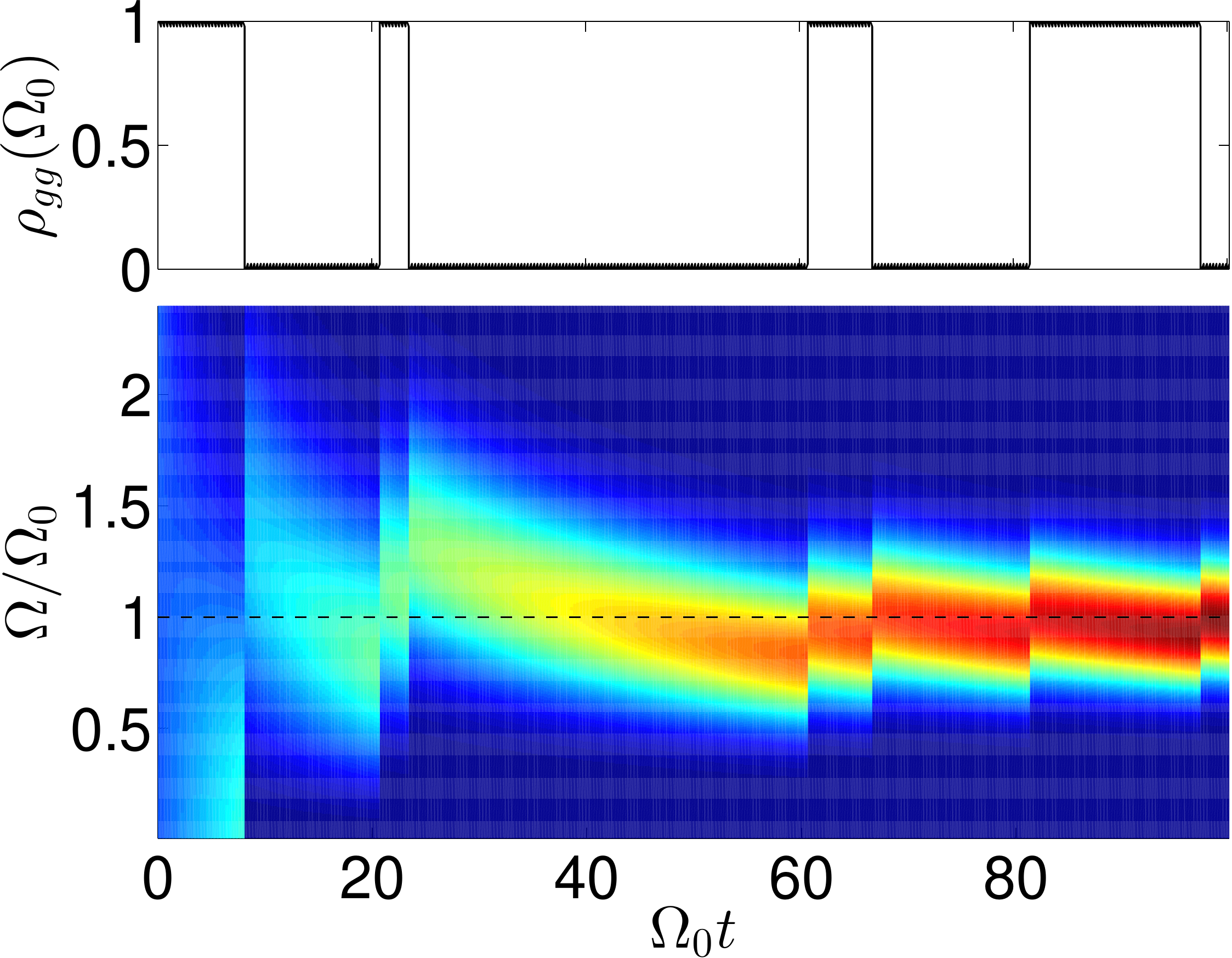}
}
%\hfill
\subfloat[\label{subfig-3:dummy}]{
\includegraphics[trim=0 0 0 0,width=0.66\columnwidth]{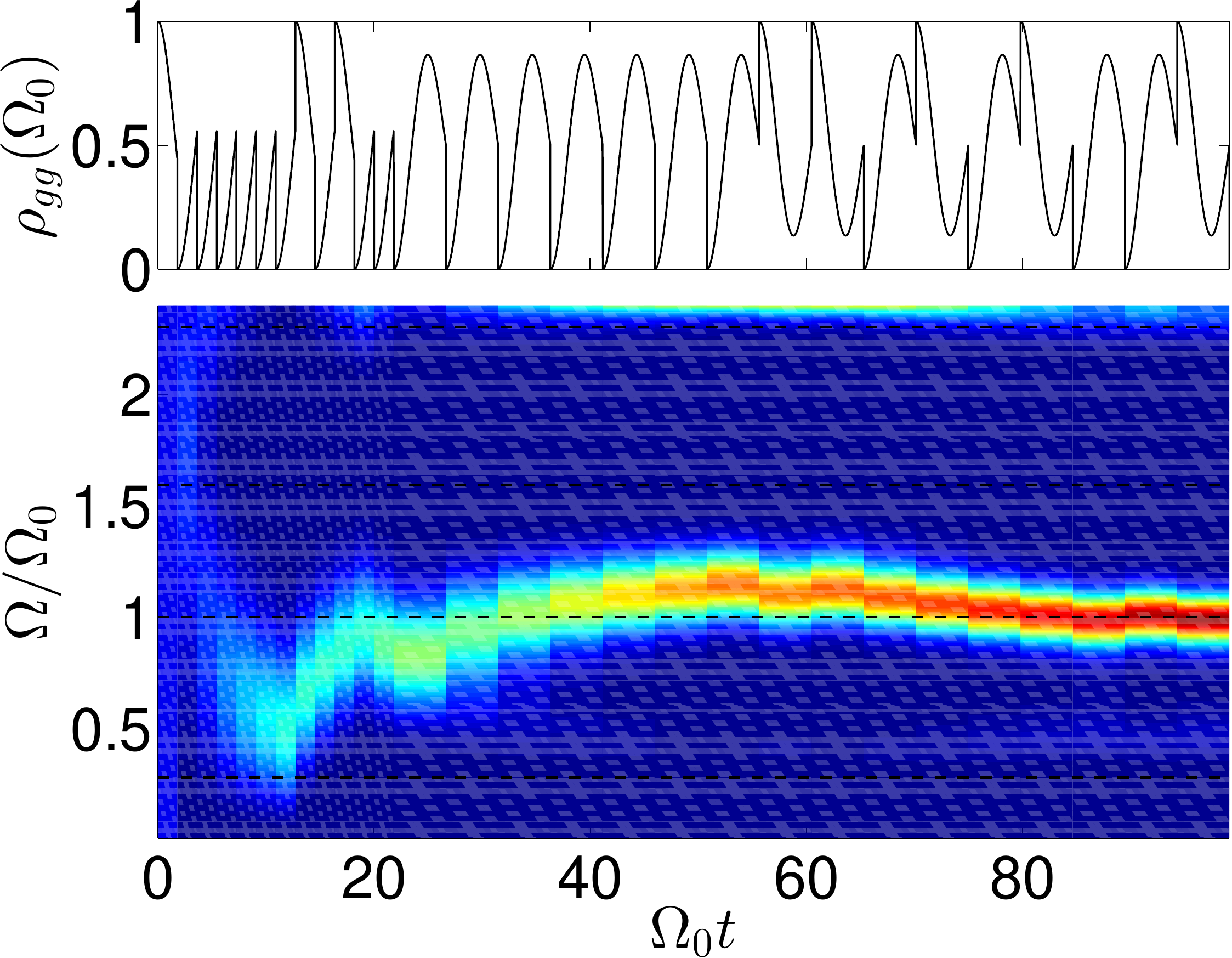}
}
\caption{\textsl{(Color online)
The upper panels show the ground state population during a simulation with $\Omega=\Omega_0$, $\delta=0$ and $\gamma = 0.1\Omega_0$.
The lower panels shows the evolution of the quasi continuous probability distribution $P(\Omega|D)$ for the Rabi frequency, conditioned on the measurement records in the upper plots. Results are shown for (a) $\Omega_0 \tau = 4.83$  and (b) $\Omega_0 \tau = 0.3$. In (c) the measurement is optimized (see text) with 12  measurements with a separation given by $\Omega_0 \tau = 1.82$ followed by 16 measurements with a separation given by $\Omega_0 \tau = 4.83$. The dashed, black lines represent values of $\Omega$ fulfilling the condition in \eqref{eq:nProblem}. Note, that $\Omega$ is assumed to lie between $0$ and $2.5\Omega_0$.
%\\Made with BayesVaryTau.m.
\label{fig:Bayes2}
}}
%\label{fig:Bayes2}
\end{figure*}

\fref{fig:Bayes2} shows the results of applying the same procedure for a wider range of candidate values of the Rabi frequency on a fine grid and for a longer time. A natural estimate for the unknown parameter $\theta$ is the value maximizing the resulting $P(\theta|D)$, and with a uniform probability distribution prior to the measurements such a maximum likelihood estimate actually saturates the CRB \eqref{eq:CRB} in the asymptotic limit of many measurements \cite{Fisher}.
To distinguish infinitesimally different hypotheses (parameter values) the probing intervals should be chosen to maximize the Fisher information, and with a finite dephasing $\gamma=0.1\Omega_0$, this happens for $\Omega_0\tau=4.83$ (See \fref{fig:FisherDephasing}). However, as seen in (a), the procedure is not able to filter away other candidate values $\Omega$ fulfilling
\begin{align}\label{eq:nProblem}
\sqrt{\Omega^2+\gamma^2}\tau = 2n\pi\pm \sqrt{\Omega_0^2+\gamma^2}\tau,
\end{align}
with $n$ an integer. If $\Omega\gg \gamma$, these Rabi frequencies lead to the same ground (excited) state populations at the time $\tau$, and thus yield identical probabilities for the measurement outcomes.

A solution to this problem is analyzed in \fref{subfig-2:dummy}, where shorter time intervals with $\Omega_0\tau = 0.3$ are used to distinguish the discrete frequencies in \eqref{eq:nProblem}.
As observed in the upper panel, however, we are here far in the Zeno regime, and even though the distribution centers around the true value in the lower panel,
it is broader than the individual peaks in (a).
\\\\
An optimized strategy applies a combination of $q$ measurements with sub-optimal evolution times to distinguish the discrete frequencies in \eqref{eq:nProblem}, i.e., to select the correct branch in \fref{subfig-2:dummy}, followed by $N-q$ measurements with the optimal evolution intervals $\tau_\mathrm{optimal}$, maximizing the Fisher information.
From \eqref{eq:nProblem} all problematic candidates are ruled out by setting $\tau_s<(2\pi-\eta)/(\sqrt{\Omega_\mathrm{max}^2+\gamma^2}+\sqrt{\Omega_0^2+\gamma^2})$, where $\Omega_\mathrm{max}$ is the maximal candidate and $\eta$ is a small positive number.
%$2\pi-\omega_0\tau_s-\eta>\omega_\mathrm{max}\tau_s$
%Now, $\Omega_0\tau$  is course not known a priori, so to include the worst case scenario one must require
% $2\pi-\eta-\Omega_\mathrm{max}\tau>\Omega_\mathrm{max}\tau$, i.e.
%\begin{align}\label{eq:tauCondition}
%\tau_1<(\pi-\eta/2)/\Omega_\mathrm{max}
%\end{align})
%$\tau_1<(2\pi-\eta)/(\Omega_\mathrm{max}+\Omega_0)$.

This procedure is equivalent to one used in state estimation \cite{GutaGill}, where an asymptotically insignificant fraction  $q=N^{1-\epsilon}$ of the measurements is applied to identify an estimate for the state within a neighbourhood of size $N^{-1/2+\epsilon}$ around the true state $\rho_0$.
The state $\rho_{\Omega}(\tau)$ at time $\tau$ corresponds to a particular value of $\Omega$, and with a distance from the true state given by
\begin{align}\label{eq:dist}
L\equiv\|\rho_\Omega(\tau_s)-\rho_{0}(\tau_s)\|_1,
\end{align}
where $\|\mu\|_1=\mathrm{Tr}(\sqrt{\mu^\dagger\mu})$,
efficient distinction of $\Omega$ from $\Omega_0$ by the first $q$ measurements, dictates a lower bound for $\epsilon$,
%\begin{align}
%q = N^{\frac{1}{2}-\frac{\ln L}{\ln N}},
%\end{align}
\begin{align}
\epsilon \geq \frac{1}{2}+\frac{\ln L}{\ln N}.
\end{align}

We pick $L$ to be small enough to discern values of the Rabi frequency that would lead to opposite measurement probabilities,
i.e., for $\Omega\gg \gamma$, we set in \eqref{eq:dist} $\sqrt{\Omega^2+\gamma^2}\tau_s=\pi-\sqrt{\Omega_0^2+\gamma^2}\tau_s$
so that
$P(g|g,\Omega)$ = $1-P(g|g,\Omega_0)$.
Note, that $L$ can be rewritten as twice the difference in ground state populations.

Finally, we obtain the optimal parameters for $N,\ q$ and $\tau_s$ numerically subject to the total time constraint
$T = q\tau_s+(N-q)\tau_\mathrm{optimal}$.

This scheme is implemented in \fref{subfig-3:dummy}. The optimized measurement parameters are given in the figure caption. The measurements in the first $t\sim 22\Omega^{-1}$ efficiently filter away the problematic candidates of \eqref{eq:nProblem}, and the subsequent optimal measurements provide a narrow peak  around the correct Rabi frequency. Hence, the hybrid method proves successful.
Note that while we presented the scheme as if the first measurements are needed to identify the approximate range of $\Omega$ first, it was recently demonstrated how adaptive measurements relying on real time Bayesian updating can be applied for optimized sensing with a single electron spin \cite{Hanson}. 

%, for our example the application of Bayes rule and the final estimate of $\Omega$ are independent of the order of the short and long measurement intervals.

\section{Conclusion and outlook} \label{sec:DISCUSS}
\begin{figure}
\centering
\includegraphics[trim=0 0 0 0,width=0.95\columnwidth]{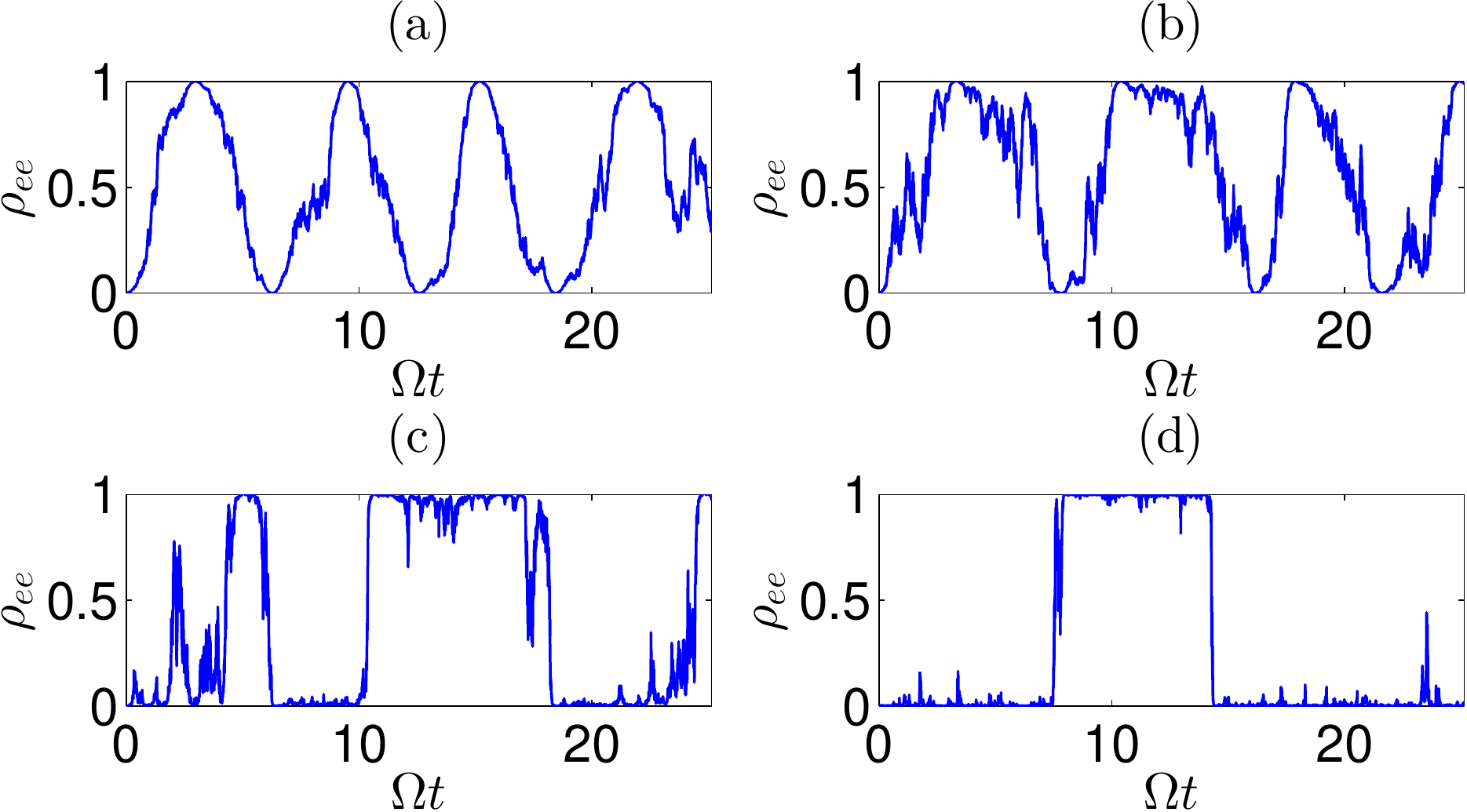}
\caption{\textsl{(Color online) Illustration of the Quantum Zeno effect for a two-level system driven on resonance subject to continuous dispersive probing with strength $\Gamma$ of the eigenstate populations. (a): $\Gamma=0.05\Omega$, (b): $\Gamma=0.2\Omega$, (c): $\Gamma=\Omega$ and (d): $\Gamma=3\Omega$. Note, the resemblance with the plots in \fref{fig:ZenoEffect}.
%Made with WeakMeasurements.m.
}}
\label{fig:ZenoEffectWeak}
\end{figure}
In this article, we have presented an analysis of the Fisher
information for the determination of parameters that govern the dynamics of
a quantum system subject to projective measurements at a given frequency.  We identified and analyzed the inhibition of precise parameter estimation caused by the Quantum Zeno effect.
%This effect is present, whenever a linear term in the survival probability of the state of a frequently probed system is absent,
Our results directly relate the Zeno inhibited evolution of a quantum system to a vanishing information acquired in any given time from the measurements causing the inhibition.
Our exemplification through the monitored Rabi oscillations of a two-level system shows that, in the Zeno limit of frequent probing, the parameter sensitivity  approaches zero, and that the interval of propagation between measurements should be made as large as possible. In the presence of dephasing, finite intervals are favourable for optimal estimation purposes, while the Zeno effect still excludes frequent probing.

%We identified the QZE as the cause of a vanishing Fisher information at $\tau\rightarrow 0 $, and
We emphasize that in the absence of a Zeno effect, where the linear term in the survival probability \eqref{eq:P0} is non-zero, the Fisher information per time may still vanish in the limit
 $\tau\rightarrow 0$ if this term is independent of $\theta $.
As an example, consider the inclusion of spontaneous decay at a rate $\gamma_{\mathrm{spont}}$ in our two-level model. With an initial excited state this yields $\Tr{\rho_0\mathcal{L}[\rho_0]}=\gamma_{\mathrm{spont}} \neq {0}$, and the Fisher information for estimating Hamiltionian parameters indeed vanishes, while the information for estimating $\gamma_{\mathrm{spont}}$ remains finite in the limit $\tau\rightarrow 0$.

 The Fisher information and the Cramér Rao bound quantify the optimal asymptotic sensitivity, but to reach unambiguous sensing at that limit, we must employ a strategy that distinguishes among remote candidate parameter values. We discussed and demonstrated a Bayesian estimation protocol, that spends a small fraction of the total probing time to establish a rough estimate, from which the asymptotic results are valid. The problem of parameter estimation from periodic probability distributions is elaborated in \cite{Ferrie2013}.

For simplicity, and for the possibility to establish analytical results, our model focused on projective (strong) measurements performed at discrete times, and we introduced the Zeno effect as the limit of very frequent measurements. We believe, however, that our results are also valid for the interpretation of different measurement scenarios, such as continuous, dispersive probing. In such experiments, a system is interrogated by a laser or microwave field, which undergoes a phase shift that depends on the state of the system, and by adjusting the intensity  of the probe field, both weak and strong (projective) measurements may be implemented on the time  scale of the system dynamics. The probe strength $\Gamma$  thus plays a role equivalent to the frequency $1/\tau$ of projective measurements in the present work, cf., the similarity of the conditional dynamics in \fref{fig:ZenoEffectWeak} for continuous probing simulated according to \cite{PhysRevA.77.052111,PhysRevA.57.1509} and \fref{fig:ZenoEffect} for the frequent projective measurements.

\section{Acknowledgements}
The authors acknowledge financial support from the Villum Foundation.

%\bibliographystyle{apsrev4-1.bst}
%\bibliographystyle{plain}

%\bibliography{master}


\end{document}